\documentstyle[aasms4]{article}

\slugcomment{Accepted for publication in \it The Astronomical Journal \rm}

\lefthead{Gibson et~al.}
\righthead{The Spectroscopic Age of 47~Tuc}

\begin{document}

\title{The Spectroscopic Age of 47~Tuc}
\author{Brad K. Gibson\altaffilmark{1,2}, 
Darren S. Madgwick\altaffilmark{2},
Lewis A. Jones\altaffilmark{3},
Gary S. Da Costa\altaffilmark{2} and 
John E. Norris\altaffilmark{2}}

\altaffiltext{1}{Center for Astrophysics \& Space Astronomy, 
Department of Astrophysical \& Planetary Sciences,
University of Colorado,
Campus Box 389, Boulder, Colorado, 80309-0389, USA}
\altaffiltext{2}{Research School of Astronomy \& Astrophysics, The
Australian National
University, Weston Creek Post Office, Weston, ACT, 2611, 
Australia}
\altaffiltext{3}{Department of Astrophysics \& Optics, School of Physics, 
University of New South Wales, Sydney, NSW, 2052, Australia}

\def\kms{km\,s$^{-1}$}
\def\spose#1{\hbox to 0pt{#1\hss}}
\def\simlt{\mathrel{\spose{\lower 3pt\hbox{$\mathchar"218$}}
     \raise 2.0pt\hbox{$\mathchar"13C$}}}
\def\simgt{\mathrel{\spose{\lower 3pt\hbox{$\mathchar"218$}}
	  \raise 2.0pt\hbox{$\mathchar"13E$}}}
\def\eg{{\rm e.g.,}}
\def\ie{{\rm i.e.,}}
\def\etal{{\rm et~al.}}
\def\aj{{AJ}}
\def\araa{{ARA\&A}}
\def\apj{{ApJ}}
\def\apjs{{ApJS}}
\def\apss{{Ap\&SS}}
\def\aap{{A\&A}}
\def\aapr{{A\&A~Rev.}}
\def\aaps{{A\&AS}}
\def\azh{{AZh}}
\def\jrasc{{JRASC}}
\def\mnras{{MNRAS}}
\def\pasa{{PASA}}
\def\pasp{{PASP}}
\def\pasj{{PASJ}}
\def\sovast{{Soviet~Ast.}}
\def\ssr{{Space~Sci.~Rev.}}
\def\zap{{ZAp}}
\def\nat{{Nature}}
\def\aplett{{Astrophys.~Lett.}}
\def\fcp{{Fund.~Cosmic~Phys.}}
\def\memsai{{Mem.~Soc.~Astron.~Italiana}}
\def\nphysa{{Nucl.~Phys.~A}}
\def\physrep{{Phys.~Rep.}}

\begin{abstract}
High signal-to-noise integrated spectra of the metal-rich globular cluster
47~Tuc, spanning the H$_{\gamma_{\rm HR}}$ and Fe4668 line indices, have been
obtained.  The combination
of these indices has been suggested (Jones \& Worthey 1995, ApJ, 446,
L31) as the best available mechanism
for cleanly separating the age-metallicity degeneracy which hampers the dating
of distant, unresolved, elliptical galaxies.  For the first time, we
apply this technique to a nearby spheroidal system, 47~Tuc, for which
independent ages, based upon more established methods, exist.  
Such an independent test of the technique's suitability has not been attempted 
before, but is an essential one before its application to more distant, 
unresolved, stellar populations can be considered valid.  
Because of its weak series
of Balmer lines, relative to model spectra, our results imply a
spectroscopic ``age'' for 47~Tuc well in excess of 20 Gyr, at odds with the
colour-magnitude diagram age of $14\pm 1$ Gyr.  The derived metal abundance,
however, is consistent with the known value.  
Emission ``fill-in'' of the
H$_\gamma$ line as the source of the discrepancy
cannot be entirely excluded by existing data, although the observational
constraints are restrictive.
\end{abstract}


\section{Introduction}
\label{introduction}

The star formation history of elliptical galaxies still remains, at many
levels, a mystery.  The fundamental uncertainty afflicting 
attempts to determine the ages of such (presumably) old stellar populations 
lies in the unfortunate fact-of-life that age and metallicity have almost
identical effects upon broad-band colours and on most line indices
(O'Connell 1976).  This age-metallicity \it degeneracy \rm (sometimes referred 
to as Worthey's
(1994) ``3/2-rule''\footnote{The 3/2-rule simply states that if two 
stellar populations have $\Delta\log({\rm Age})/\Delta\log({\rm 
Metallicity})\equiv -3/2$, their broad-band colours and (most)
line indices will be nearly identical.})
has led many (\eg\ O'Connell 1976; Worthey 1994; Rose 1994) to
search for individual absorption features which may be
sensitive to either age \it or \rm metallicity, but not both.  

Unfortunately, line indices have not proven to be the panacea once hoped for.
As Worthey (1994) has shown (see his Table 6), the vast majority of 
low-resolution lines defined by the Lick IDS survey (\eg\ Worthey \etal\ 
1994) lie uncomfortably close to the 3/2-rule degeneracy ``zone''.  An
encouraging result to come out of the Worthey (1994) IDS study, however,
was the
identification of one particularly useful, and strong, 
indicator, apparently free of any strong age influence.  Fe4668, whose bandpass
spans $\sim 4635\rightarrow 4722$\,\AA, tracks not only Fe, but contains a 
strong
contribution from C$_2$, Ti, Cr, and Mg, and as Worthey, Trager \& Faber (1996)
note, is the best current candidate tracer for the mean metallicity ``Z'',
almost exclusively sensitive to this alone.

On the other hand, identifying a matching ``pure'' age discriminator has proven
more difficult.  H$_\alpha$ and H$_\beta$ have long been popular indices, in 
this regard (\eg\ O'Connell 1976), but as stressed by Gonz\'alez
(1993), both lines can be easily distorted (even filled in) by emission from
ionized gas, \it severely \rm
compromising their usage.  Higher-order Balmer emission
lines (\eg\ H$_\delta$ and H$_\gamma$)
are admittedly weaker, yet free of such ``emission'' complications, and appear
to provide a very sensitive age diagnostic.  Rose (1994) recognized this latter
fact, and designed a set of higher-resolution line indices
in order to take advantage of these age discriminating
``powers''.  The Rose system employs a 3.74\AA\ window over which the line
indices are measured; in contrast, the Lick IDS system (typically) uses an
$\sim$40\AA\ window.  One index in particular,
H$_{\gamma_{\rm HR}}$ (centred at 4340.5\,\AA), was strongly sensitive to age
alone.\footnote{The ``HR'' subscript stands for ``high resolution'', in order
to distinguish the Rose-style equivalent widths, with 3.74\,\AA\, bandpasses,
from the Lick-style indices which are an order of magnitude broader.}

The near-orthogonal behaviour, in the age-metallicity plane,
of the H$_{\gamma_{\rm HR}}$ and Fe4668 line
indices, was demonstrated by Jones \& Worthey (1995; Figure 2), and suggests
a viable mechanism for breaking the aforementioned age-metallicity
degeneracy.  An extension of this Lick+Rose index comparison was 
undertaken by Jones (1995,1996ab), but now concentrating solely upon pairs
of high resolution Rose indices -- \eg\
Jones examined the H$_{\gamma_{\rm HR}}$--Ca\,I$_{\rm HR}$ age-metallicity 
diagnostic diagram.

Jones \& Worthey (1995) stress several caveats regarding their
proposed optimal age-dating technique.  First, while the parameter-space
coverage (\eg\ T$_{\rm eff}$, $\log g$, and [Fe/H]) of their stellar library,
upon which the model sequences are ultimately based, is good, it is
not complete.  Specifically, there is an admitted lack of metal-poor stars,
especially hot horizontal branch stars, which could contribute to the Balmer
line absorption.  Because of this,
Jones \& Worthey emphasize that their sequences are not directly
applicable to systems with metallicities outside $-1.0\le{\rm [Fe/H]}\le +0.6$.
Second, in order to apply this technique accurately 
to integrated spectra of spheroidal systems, very high signal-to-noise ratios
($\simgt 100$), at moderately high spectral resolutions ($\sim 1$\,\AA), are a
necessity.

To this end, Jones \& Worthey (1995) present such a spectrum for M32,
and find an age of $\sim 6\rightarrow 7$ Gyr, in agreement with
O'Connell (1980) and Rose (1994).  While this result is a necessary proof
of the technique, it is not a sufficient one, as it effectively compares
two population \it
synthesis \rm methods, both of which are \it indirectly \rm determining
an age estimate.  Optimally, any testing of a new, and
potentially exciting, technique should be completely ``independent'', at some
level.  Unfortunately, there are simply no independently age-dated ellipticals
known.  
On the other hand, there does exist another family of systems which are
ideally suited to such a study -- the metal-rich globular clusters of the Milky
Way. \it
Metal-rich globular clusters are the only practical systems with which to 
independently assess 
the reliability of the H$_{\gamma_{\rm HR}}$--Fe4668 age-dating technique. \rm
This comes about because 
accurate ages have already been determined for at least four
clusters in the metal-rich regime to which the Jones \& Worthey technique is
applicable (\eg\ Richer \etal\ 1996).
With its high metallicity (\ie\
[Fe/H]$\approx -0.7$) and colour-magnitude diagram (CMD)-derived 
age of $14\pm 1$ Gyr
(Richer \etal\ 1996), the southern globular cluster, 47~Tuc, is an ideal
laboratory within which to test this new age-dating technique.

Existing high-resolution integrated spectra of 47~Tuc in the literature
include those of Smith (1979) and Rose (1994).
In both cases though, the observations were
restricted to $\lambda<4400$\,\AA, meaning the metallicity-discriminant
Fe4668 line index could not be measured.

In Section
\ref{observations}, we briefly describe our new observations of 47~Tuc,
which now provide simultaneous coverage of both the
age-sensitive feature 
H$_{\gamma_{\rm HR}}$, and the
metallicity-sensitive features Ca\,I$_{\rm HR}$ and Fe4668.
Such data will allow us to test how
accurately the spectroscopic age of this simple stellar population agrees
with its accurately determined CMD age. 
The results of our analysis are
discussed in Section \ref{discussion}, and summarised in Section \ref{summary}.

\section{Observations and Analysis}
\label{observations}

Two high signal-to-noise ($S/N\simgt 100$) integrated spectra of 
47~Tuc were acquired during the nights of October 20 and 21,
1996, using 
the Coud\'e spectrograph, with the 32$^{\prime\prime}$ Schmidt camera
(Grating E; 1$^{\rm st}$ order), 
on the 74$^{\prime\prime}$ telescope at Mount Stromlo Observatory,
was employed.  
The $1^\prime.1$ long slit was 
scanned back and forth (E-W) across the slit, to limits of $\pm 0^\prime.55$
from the cluster center (total coverage 
in excess of 2.5 core radii, ensuring stochastic
fluctuations due to finite sampling were minimized), for
$\sim 30$ minutes per exposure.  
The 24 $\mu$m pixels of the
Tek 2k$\times$2k CCD translated to 0.48\,\AA\, per pixel -- \ie\ 983\,\AA\, of
spectral coverage, which for a central wavelength setting of $\sim 4500$\,\AA,
allowed simultaneous determination of 
Ca I$_{\rm HR}$, H$_{\gamma_{\rm HR}}$, and Fe4668 indices.  Companion
blank-sky exposures (again, of $\sim 30$ minutes duration,
taken $\sim 20^\prime$ to the north of
the cluster core) were acquired subsequent to each 47~Tuc exposure. The
relative strength of the sky versus cluster signal was negligible.
For calibration purposes, 13 stars (spanning
a range in metallicity and spectral type) in common with
Jones' Coud\'e Feed Spectral
Library (see Section 2.4 of Leitherer \etal\ 1996 and Jones 1996b) 
and the Lick/IDS Library
(Worthey \etal\ 1994) were observed, as were two
spectrophotometric standards.  Standard IRAF tasks were employed for
the flat-fielding, flux, and wavelength calibration.  Observations taken
throughout the run of bright, metal-poor halo giants imply that sky
subtraction has not introduced any unforeseen spectral artifacts.
Because the central one-dimensional velocity dispersion of 47~Tuc is only
$\sigma_c=11.6\pm 1.4$\,km\,s$^{-1}$ (Meylan \& Mayor 1986), no correction for
internal dispersion was applied to the spectra; as Trager \etal\ (1998; Figure
3) demonstrate, such a correction, for globular clusters, is 
negligible.  

Before measuring the H$_{\gamma_{\rm HR}}$ and Ca\,I$_{\rm HR}$ line indices,
our observed standards
and target 47~Tuc spectra were first smoothed 
with a $\sigma=103$\,km\,s$^{-1}$
Gaussian, and re-binned to 0.623\AA/pixel, in both cases
to parallel Jones (1996b).
The H$_{\gamma_{\rm HR}}$ and Ca\,I$_{\rm HR}$ equivalent widths are measured
as detailed in Section 3 of Jones \& Worthey (1995).  Flanking blue and red
pseudo-continuum peaks, usually associated with the immediately-adjacent blue
and red shoulders of the relevant absorption line, were identified; the
equivalent widths were then measured relative to the continuum formed by
connecting the two pseudo-continuum peaks.  For H$_{\gamma_{\rm HR}}$, the blue
pseudo-continuum was associated with the local maximum between 4329\,\AA\,
and 4336\,\AA, while the red pseudo-continuum was associated with the maximum
in the 4346$-$4350\,\AA\, 
window; for Ca\,I$_{\rm HR}$, the correspondings windows
were 4218$-$4222\,\AA\, and, either 4230$-$4232\,\AA\, (early spectral types)
or 4242$-$4246\,\AA\, (later spectral types).
Figure \ref{fig:fig1} shows the pair of integrated 47~Tuc spectra in the
region of both the H$_{\gamma_{\rm HR}}$ and Ca\,I$_{\rm HR}$ indices; the
associated flanking pseudo-continua are similarly noted.  
The equivalent width was then derived via a simple Simpson's Rule algorithm,
summing the area associated with the seven 0.623\,\AA\, pixels
centred on the minimum of the high resolution absorption 
feature\footnote{A quirk of the line-measuring code employed by Jones 
(1995,1996ab), itself an offspring of that used by Rose (1994, and earlier 
papers), is that
the equivalent width derivation requires three pixels on either side of
absorption feature minimum (itself one pixel) -- \ie\ while the pixel
centre-to-centre window is the 3.74\,\AA\ referred to throughout this
paper (and earlier Rose and Jones papers), the effective area over which the
integration is acting corresponds to seven pixels, or 4.36\,\AA.}
-- Gaussian-fitting to the line profiles was not employed in our analysis, 
again, in order to maintain consistency with Jones (1996b). 
The derived
equivalent widths of Figure \ref{fig:fig1} are
consistent, from night-to-night, at the $2\rightarrow 3$\% level.

\placefigure{fig:fig1}

For the Fe4668 measurement, the 13 standards and 47~Tuc spectra were smoothed
with a $\sigma=250$\,km\,s$^{-1}$ Gaussian, to match the Lick IDS resolution.
Adjacent blue and red continua were determined from the mean flux in the 
4611.50$-$4630.25\,\AA\, and 4742.75$-$4756.500\,\AA\, windows, respectively.
Using the continuum defined by the line connecting the mean of these two
windows, the
equivalent width was then derived by integrating the absorption 
feature over the 4634.00$-$4720.25\,\AA\, window.  The adopted windows are
marginally different from those defined by Worthey \etal\ (1994; Table 1), but
reflect those adopted in the 1997 edition of the Lick/IDS stellar catalog
(Trager \etal\ 1998).

Comparison of the equivalent widths derived from our 13 standards with those 
tabulated in
the Coud\'e Feed Spectral Library (H$_{\gamma_{\rm HR}}$ and Ca\,I$_{\rm HR}$)
and Lick/IDS Library (Fe4668) revealed the two systems were on the
scale, and no further corrections were 
needed to bring the systems into agreement -- \ie\ both groups were measuring
identical equivalent widths for the same standards -- and therefore, our
measured 47~Tuc equivalent widths could be used \it as is \rm in comparing
against the model age-metallicity diagnostic diagrams of Jones \& Worthey
(1995) and Jones (1996a).  Evidence to this effect is presented in
Figure \ref{fig:fig2}, in which our standard star Ca\,I$_{\rm HR}$ 
equivalent widths (x-axis) are plotted against those from Jones' (1996b)
Coud\'e Feed Spectral Library (y-axis) - the mean offset (in the sense of
Jones$-$Present Study) is only $+0.010\pm 0.006$\AA, a factor of
$\sim$3$\times$ smaller than the typical uncertainty associated with
pseudo-continua placement and night-to-night variations (recall
Figure \ref{fig:fig1}).  Identical conclusions hold for the other line indices
in our study.
Our measured (and adopted)
47~Tuc equivalent widths are therefore
$1.032\pm 0.040$\,\AA\, (H$_{\gamma_{\rm HR}}$), 
$1.042\pm 0.030$\,\AA\, (Ca\,I$_{\rm HR}$), and
$1.95\pm 0.45$\,\AA\, (Fe4668), where the uncertainties reflect those due to
pseudo-continuum placement, and, to a lesser degree, the scatter derived from
repeat measurements of the set of standards.

\placefigure{fig:fig2}

\section{Discussion}
\label{discussion}

Following the prescription outlined by Jones \& Worthey (1995) and
Jones (1996b), the Worthey (1994)
population synthesis models were used to predict H$_{\gamma_{\rm HR}}$,
Ca\,I$_{\rm HR}$, and Fe4668 line index strengths, as a function of age and
metallicity [Fe/H], for a grid of single burst stellar populations.
Twenty-four realizations, spanning ages $2\rightarrow
17$\,Gyr and metallicities [Fe/H]=$-0.50\rightarrow +0.50$, were made; the full
set of 24 models were then incorporated into the H$_{\gamma_{\rm
HR}}$--Ca\,I$_{\rm HR}$ age-metallicity diagnostic grid shown in Figure
\ref{fig:fig3}.  While not perfectly orthogonal, this pair of high resolution
Rose-style indices do provide adequate separation of age and
metallicity, with a minimum of degeneracy.  The four filled symbols represent
the same set of low luminosity ellipticals shown in the earlier analysis of
Jones (1996a), but now smoothed to the same $\sigma=103$\,km\,s$^{-1}$
resolution adopted by Jones (1996b) and the present study.  As discussed
previously (Jones 1996a), these four galaxies represent a (luminosity-weighted)
solar metallicity age sequence spanning the range $2\rightarrow 12$\,Gyr.

\placefigure{fig:fig3}

The open starred
symbol represents our new 47~Tuc data point; unlike the case for the
galaxies shown in Figure \ref{fig:fig3}, 47~Tuc has an independent 
age\footnote{Salaris \& Weiss (1998), using $\alpha$-element enhanced stellar
models, have recently revised the age of 47~Tuc from the canonical 
(scaled-solar abundance) Richer \etal\
(1996) value of $14\pm 1$\,Gyr, to the lower value of $9\pm 1$\,Gyr.  On the
surface this only worsens the spectroscopic versus CMD age discrepancy of
Figure \ref{fig:fig3}, although the comparison is perhaps not a fair one, since
Worthey's (1994) synthesis models are (necessarily) tied to models and stellar
data which are likewise of scaled-solar abundances.  Similarly, the 
post-Hipparcos globular cluster distance
scale appears to favor a 2-3\,Gyr reduction
in the mean age of the oldest globulars (Chanoyer \etal\ 1998); again, this
only strengthens our conclusion regarding the spectroscopic age of 47~Tuc,
although to be fair, as before, we retain the Richer \etal\ (1996) result
as the most representative one against which to compare our results.}
($14\pm 1$\,Gyr) and metallicity ([Fe/H]$\approx -0.7$) determination, derived
from isochrone fitting to the CMD (Richer \etal\ 1996; Salaris \& Weiss 1998).
It should be immediately clear that the spectroscopic age of 47~Tuc, as
determined by the age-sensitive H$_{\gamma_{\rm HR}}$ line index is
inconsistent with its CMD age, at the $\sim$4$\sigma$ 
level, being $\sim$0.2\,\AA\, lower than expected for this 14\,Gyr old simple
stellar population.  This $\sim$0.2\,\AA\, H$_\gamma$ discrepancy was
previously noted by Rose (1994; Section 5.1), although not in the context of
age-dating its integrated spectrum.  On the other hand, 
it is somewhat re-assuring that the
metallicity inferred from Figure \ref{fig:fig3} (\ie\ [Fe/H]$<$-0.5) is
consistent with that derived from the CMD (Richer \etal\ 1996).

Figure \ref{fig:fig4} shows the 
H$_{\gamma_{\rm HR}}$--Fe4668 age-metallicity diagnostic diagram as 
presented by Jones \& Worthey (1995; Figure 2);  the same nineteen models
used in their analysis are incorporated in our grid.  The only (subtle)
difference between Figure \ref{fig:fig4}, and that of Jones \& Worthey,
is that, in the latter, the spectra were smoothed to 
$\sigma=80$\,km\,s$^{-1}$ (to match the resolution of their M32 spectrum),
while in the present study, recall that
 a $\sigma=103$\,km\,s$^{-1}$ Gaussian was
employed (to match the spectra and models of Jones 1996b).  Again, 
the model sequences of Figure \ref{fig:fig4} clearly demonstrate the
near perfect orthogonal nature of the H$_{\gamma_{\rm HR}}$ and Fe4668
line indices in the age-metallicity plane.

\placefigure{fig:fig4}

As in Figure \ref{fig:fig3}, the filled square represents the Jones \&
Worthey (1995) observation of M32, while the open starred symbol shows our
new 47~Tuc data point.  As was the case for the pair of Rose-style 
high resolution line indices (Figure \ref{fig:fig3}), in the 
H$_{\gamma_{\rm HR}}$ versus Fe4668, the anomalously low 
H$_{\gamma_{\rm HR}}$ in the integrated spectrum of 47~Tuc results in a
spectroscopic age ($\simgt$20\,Gyr) which is clearly discrepant with
CMD age ($14\pm 1$\,Gyr).

Is there some way in which to reconcile the spectroscopic versus CMD
age discrepancy?  As discussed by Rose (1994), virtually every line index
in the integrated spectrum of 47~Tuc appears perfectly well-behaved,
except the anomalously strong CN bands and anomalously weak H$_\gamma$.
Jones (1996b) speculated that perhaps the H$_\gamma$ ``problem'' was
being caused by CH contamination in the H$_\gamma$ line; because the
central regions of
47~Tuc (\ie\ the region over which our integrated spectrum
was obtained) is replete with CN-strong giants (Norris \& Freeman 1979;
Da Costa 1997),
and there exists the well-known CN-CH anticorrelation for CN-strong
giants (Norris, Freeman \& Da Costa 1984), a scenario was envisaged whereby
H$_\gamma$ might be weaker in CN-strong (CH-weak) stars.  
Jones (1996b), though,
was able to show, using a sample of 11 CN-strong and CN-weak
giant and horizontal branch stars from M71 (a cluster similar to 47~Tuc),
that this initial hypothesis was not supported by the data.


The most obvious speculative source of the discrepancy is that some 
unidentified source of
H$_\gamma$ \it emission \rm is acting to ``fill in'' the
H$_\gamma$ \it absorption \rm feature.  Hesser \& Shawl (1977) provide an upper
limit to the emission flux at H$_\alpha$ of F(H$_\alpha$)$\simlt 1.3\times
10^{-11}$\,erg\,cm$^{-2}$\,s$^{-1}$.  Here we have used their most restrictive
dataset - \ie\ $C_1=26.3$\,s$^{-1}$ (from their 22/08/75 observations), coupled
with the definition F(H$_\alpha$)$\approx 5\times
10^{-13}C_1$\,erg\,cm$^{-2}$\,s$^{-1}$, from Smith, Hesser \& Shawl (1976).
Now, the core of 47~Tuc, with M$_V\approx +6$ (and ignoring reddening), 
corresponds to a flux of 
$\sim 1.6\times 10^{-11}$\,erg\,cm$^{-2}$\,s$^{-1}$\,\AA$^{-1}$.  So, the
Hesser \& Shawl upper limit of $1.3\times 10^{-11}$\,erg\,cm$^{-2}$\,s$^{-1}$
corresponds to $1.3/1.6=0.8$\,\AA\ of equivalent width at H$_\alpha$.
Several simplifying assumptions must be made to go from H$_\alpha$ to
H$_\gamma$ - specifically, 
under case B conditions and a temperature of $10000$\,K,
radiative recombination leads to an intrinsic ratio 
L(H$_\alpha$)/L(H$_\gamma$)$\approx 6.5$, which gets reduced by a factor 1.7,
reflecting the ratio of continuum flux at H$_\alpha$ to that at H$_\gamma$
(employing the Worthey 1994 models).  In other words, H$_\gamma$ is a factor of
$3.8\times$ weaker than H$_\alpha$; recalling the constraint at H$_\alpha$ of
$\simlt 
0.8$\,\AA, this corresponds to $\simlt 
0.21$\,\AA\ of equivalent width at H$_\gamma$.

From Figures \ref{fig:fig3} and \ref{fig:fig4}, we can see that the 
H$_\gamma$ ``discrepancy'' amounts to $\sim 0.17$\,\AA, in that we \it a priori
\rm expected an equivalent width of $\sim 1.20$\,\AA, to match the known CMD
age of 47~Tuc, as opposed to the $\sim 1.03$\,\AA\ observed.  Recalling that
the inferred upper limit to emission fill-in at H$_\gamma$ (from the previous
paragraph) was $\simlt 0.21$\,\AA, we can see that the requisite $\sim
0.17$\,\AA\ of emission needed to reconcile the CMD and spectroscopic ages is
not entirely excluded by the Hesser \& Shawl (1977) observations, although 
little room is left to maneuver should new H$_\alpha$ emission observations
push this limit even lower.

The situation regarding the anomalously low H$_\gamma$ index in 47~Tuc
remains as it was at the time of the Rose (1994) and Jones (1996b) studies - 
\ie\ 47~Tuc appears normal in virtually every line index, except H$_\gamma$,
and to date no satisfactory explanation has been put forth.  Emission fill-in
has not been entirely excluded, but the constraints are tight, as shown above.
Should such emission prove to be the ultimate resolution of the discrepancy
though, it will beg the question as to exactly which mechanism is responsible,
and why it differs from the local calibrating stars used to derive the model
sequences of Figures \ref{fig:fig3} and \ref{fig:fig4}.  


\section{Summary}
\label{summary}

From the outset, our goal was a simple one - to provide an independent
test of the Jones \& Worthey (1995) H$_{\gamma_{\rm HR}}$--
Fe4668 diagnostic (and the Jones 1996a
H$_{\gamma_{\rm HR}}$--Ca\,I$_{\gamma_{\rm HR}}$ diagnostic) as a tool for
breaking the long-standing age-metallicity degeneracy plaguing integrated
spectra of stellar populations.  Such a test can only be done
(properly) upon systems with independently-derived ages and metallicities;
local metal-rich globular clusters with accurate, deep, CMDs,
provide an ideal test-bed.
Our new high $S/N$, high resolution, integrated spectrum of 47~Tuc shows
anomalously low H$_\gamma$ absorption, which when tied to the diagnostic
models of Jones \& Worthey (1995), shows, for the first time, that
the inferred spectroscopic age is well in excess
of 20\,Gyr. This is
clearly at odds with the CMD age of $14\pm 1$\,Gyr.  No suitable
explanation as to the source of this discrepancy has been found, despite
the best efforts of Rose (1994) and Jones (1996b).  Emission, amounting to
$\sim 0.17$\,\AA of equivalent width at H$_\gamma$, is required to reconcile
the discrepancy.  Such emission is not excluded by the Hesser \& Shawl (1977)
observations of 47~Tuc 
(which set an upper limit of $\sim 0.21$\,\AA\ at H$_\gamma$), but
if it is the responsible mechanism, must be lurking just marginally
below the limit set by these early Fabry-Perot observations.  Clearly, a deeper
re-visit is called for.

It is true that the Jones \& Worthey (1995) 
H$_{\gamma_{\rm HR}}$--Fe4668 age-metallicity diagnostic will remain an
invaluable tool for age-dating integrated spectra of early-type systems, but
perhaps the most important conclusion to take from our simple test of the
technique is that it is not infallible.
Clearly, it will be imperative to extend our analysis to a larger sample of
metal-rich globulars, Galactic open clusters, and star clusters in the
Magellanic Clouds -- such an analysis will
demonstrate whether 47~Tuc is a pathological case or 
the inconsistent spectroscopic age derived here is indicative of 
an endemic, and currently underappreciated, flaw in the technique.

\acknowledgments

We wish to thank Guy Worthey, Jim Rose, and Ken Freeman for their helpful
advice.  A special word of thanks is aimed at the referee, who went above and
beyond the call of duty in aiding us in the interpretation of the Hesser \&
Shawl (1977) observations.
DSM acknowledges the financial support provided 
through the ANU/MSSSO Summer Research Fellowship Program.

\clearpage

\figcaption[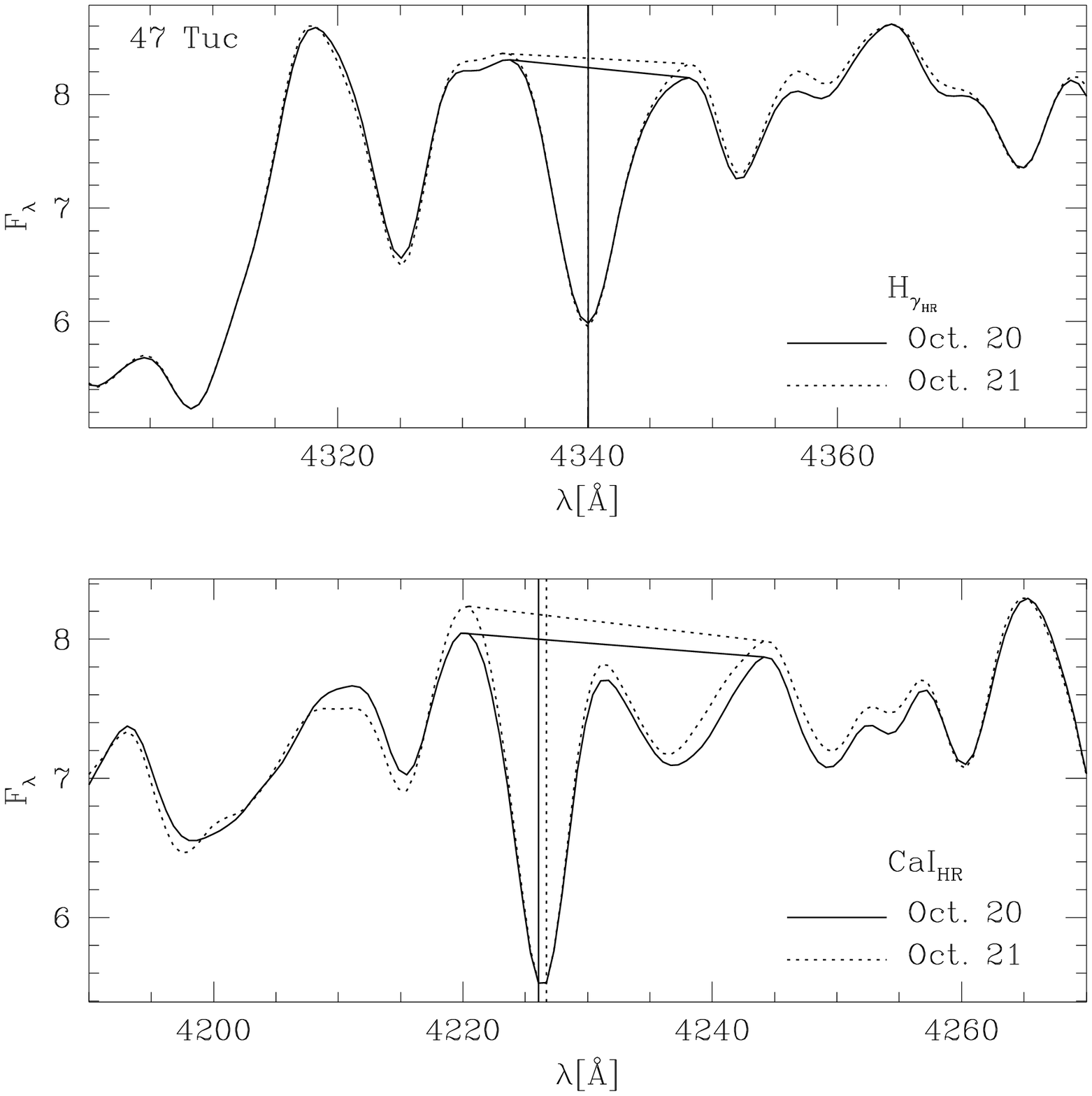]{
Integrated 47~Tuc spectra, smoothed to a velocity resolution of
$\sigma=103$\kms to match Jones (1996b), acquired the nights of October 20 
(solid) and October 21 (dotted).  The primary high-resolution line indices
employed in our study are shown - H$_{\gamma_{\rm
HR}}$ (upper panel) and Ca\,I$_{\rm HR}$ (lower panel) - as are the 
adopted neighboring pseudo-continua.  The derived equivalent widths are
consistent, from night-to-night, at the $2\rightarrow 3$\% level.
\label{fig:fig1}}

\figcaption[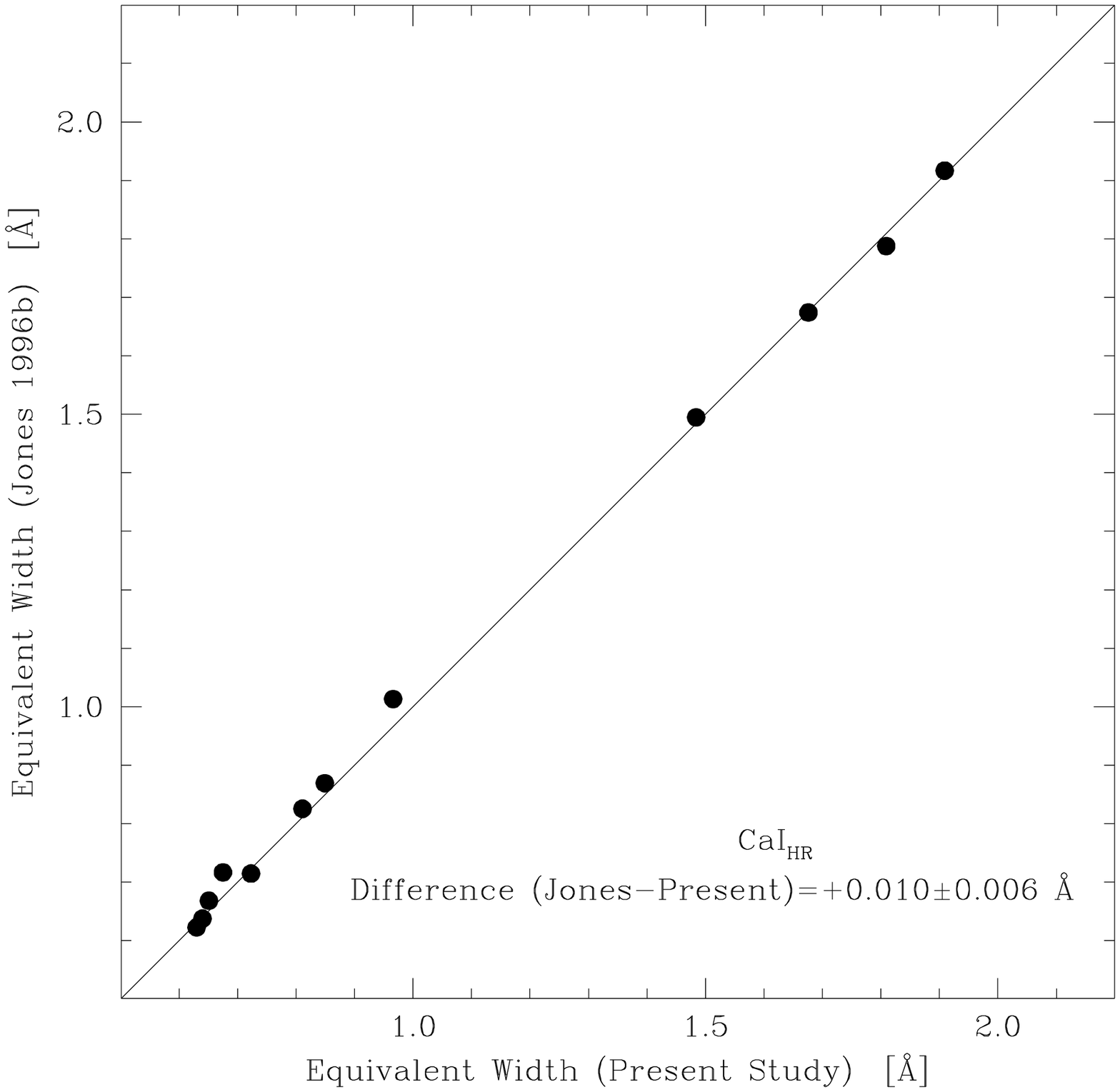]{Comparison of published Ca\,I$_{\rm HR}$
equivalent widths from the Jones (1996b)
Coud\'e Feed Spectral Library (y-axis) versus those of the current study
(x-axis), for the standard stars in common.  Clearly the two systems are
directly comparable, with the mean offset between the two samples being
only $+0.010\pm 0.006$\AA, in the sense of Jones$-$Present Study.
\label{fig:fig2}}

\figcaption[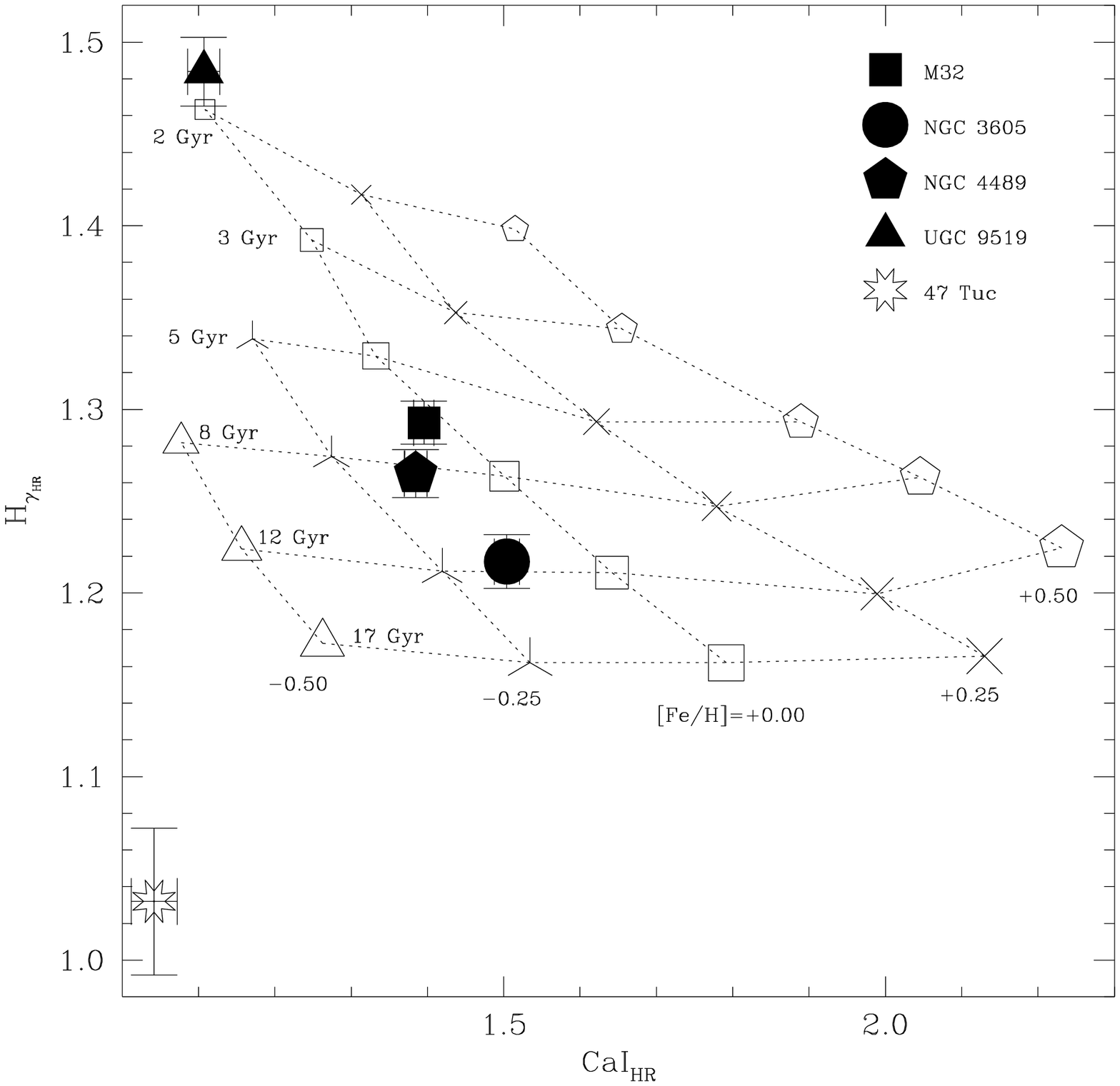]{The Rose-style age indicator 
H$_{\gamma_{\rm HR}}$ versus the Rose-style metallicity indicator Ca\,I$_{\rm
HR}$.  The open symbols represent the model sequence, increasing in size with
age, and shape with metallicity.  The solid symbols correspond to the low
luminosity ellipticals reported by Jones (1996a), but now smoothed to 
$\sigma=103$\,km\,s$^{-1}$, to match the model grid.
The predicted age ($\simgt 20$ Gyr) for the globular cluster reported 
here, 47~Tuc (open symbol), 
is clearly at odds with its CMD-age ($14\pm 1$ Gyr).
\label{fig:fig3}}

\figcaption[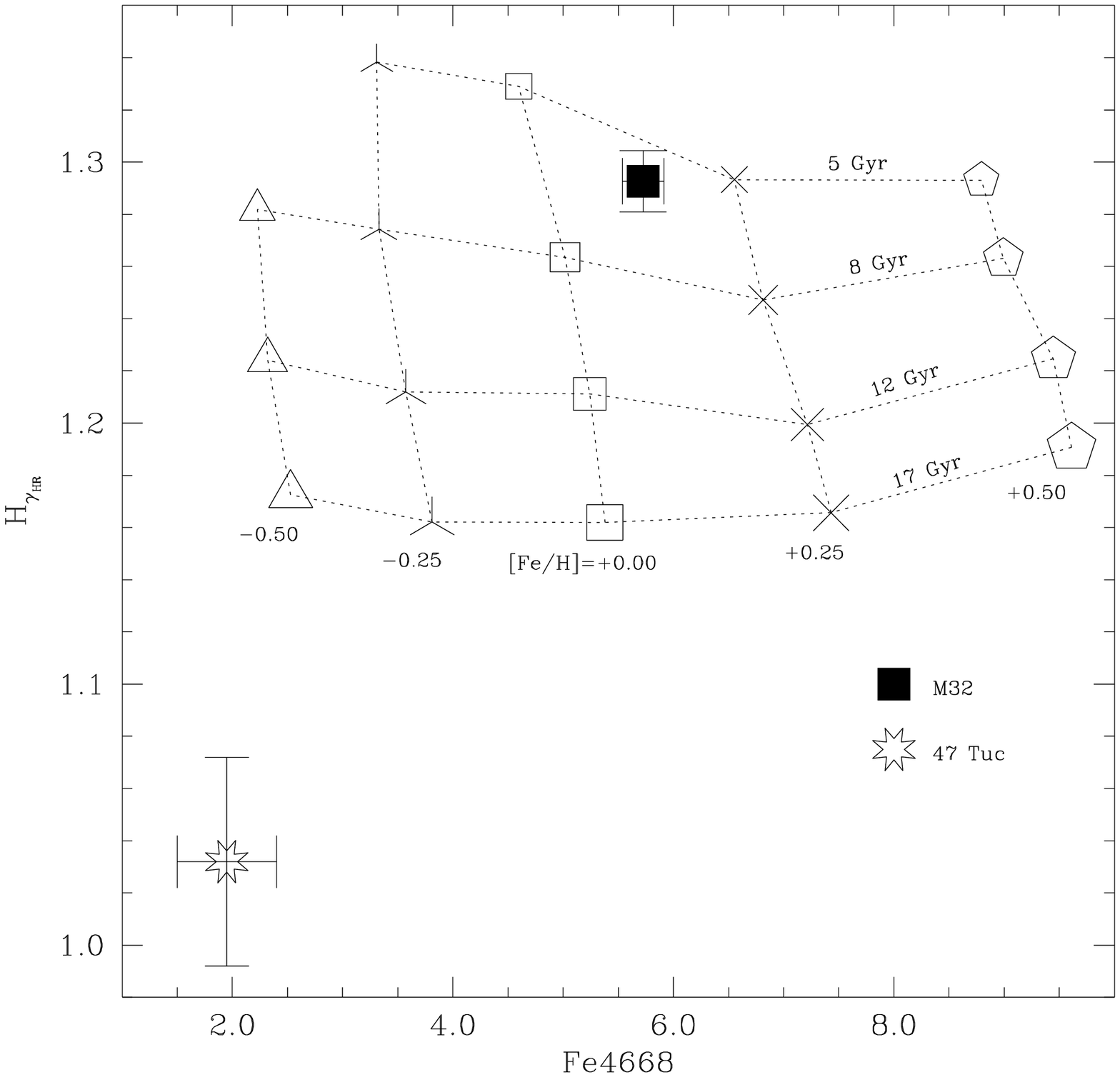]{The Rose-style age indicator
H$_{\gamma_{\rm HR}}$ versus the Lick metallicity indicator Fe4668.
The open symbols represent the model sequence, increasing in size with
age, and shape with metallicity.  The filled square corresponds to the M32
result reported by Jones \& Worthey (1995), but now smoothed,
respectively, to 
$\sigma=250$\,km\,s$^{-1}$ (for measuring Fe4668) 
and $\sigma=103$\,km\,s$^{-1}$ (for measuring H$_{\gamma_{\rm HR}}$).
As for Figure \ref{fig:fig3}, the predicted age ($\simgt 20$ Gyr) for 47~Tuc
(open symbol), is at odds with its CMD-age ($14\pm 1$ Gyr).
\label{fig:fig4}}

\clearpage

\epsscale{1.0}
\plotone{Gibson.fig1.eps}

\clearpage

\epsscale{1.0}
\plotone{Gibson.fig2.eps}

\clearpage

\epsscale{1.0}
\plotone{Gibson.fig3.eps}

\clearpage

\epsscale{1.0}
\plotone{Gibson.fig4.eps}

\end{document}